\documentclass[apjl, twocolappendix]{emulateapj}
\usepackage[fleqn]{amsmath} 
\usepackage{amsfonts} 
\usepackage{amstext}
\usepackage{amssymb} 
\usepackage[colorlinks=true, citecolor=blue, linkcolor=blue,
breaklinks=true]{hyperref}
\usepackage{natbib}
\usepackage{graphicx}

\def\etal{\it et al. \rm }

\begin{document}

\title{On the Dichotomy between Normal and Dwarf Ellipticals}
\author{James M. Schombert$^1$}
\affil{$^1$Department of Physics, University of Oregon, Eugene, OR USA 97403}

\begin{abstract}

\noindent Using images from the SDSS DR13 library, we examine the structural
properties of 374 bright (classed E0 to E6) and dwarf ellipticals (classed dE(nN) to
dE(N)).  The sample combines a multicolor sample of bright ellipticals (252 galaxies
with $M_g < -20$) with a new sample of faint ellipticals (60 galaxies with $M_g >
-20$) which overlaps the dwarf elliptical sample (62 galaxies) in luminosity and
size.  The faint ellipticals extend the linear structural correlations found for
bright ellipticals into parameter space not occupied by dwarf ellipticals indicating
a dichotomy exists between the two types.  In particular, many faint ellipticals have
significantly higher effective surface brightnesses compared to dE's which eliminates
any connection at a set stellar mass.  Template analysis of the three subsets of
ellipticals demonstrates that the bright and faint ellipticals follow the same trends
of profile shape (weak homology), but that dwarf ellipticals form a separate and
distinct structural class with lower central surface brightnesses and extended
isophotal radii.

\end{abstract}

\section{Introduction}

Their simple morphology and smooth light distribution has made elliptical galaxies
the default standard for studying galaxy structure and kinematics.   Distinguished in
their morphology only by their outer axial ratio, the class of ellipticals display
remarkable uniformity in structure parameters as a function of stellar luminosity (a
proxy for total stellar mass, see Schombert 2013).  For example, ellipticals display
a smooth correlation between characteristic scalelength and luminosity, a linear
transition in surface brightness profile shape with luminosity and bright ellipticals
have higher central stellar densities (i.e., surface brightnesses) with higher
luminosities which decreases in a uniform fashion down to total luminosities around
$M_B = -18$.

However, deep studies of the Virgo cluster suggested a different type of elliptical
exists below $M_B = -18$, the subclass of dwarf ellipticals (dE, Sandage \& Binggeli 1984).  Dwarf
ellipticals appeared to be morphologically distinct from normal ellipticals (normal
defined as ellipticals with power-lawed profiles, near $r^{1/4}$ in shape) and were
considered for many years to constituted a separate type of elliptical (Wirth \&
Gallagher 1984) by morphologists who claim there is no difficulty in separating
normal ellipticals from dwarf ellipticals simply based on their visual appearance (in
particular, their diffuseness).  Whether this perceived difference can be mapped into
quantitative structure parameters was unclear and highly debated (see Graham 2013 for
a review), but there is no doubt that bright ellipticals are power-law (i.e.,
r$^{1/4}$) in shape and dwarf ellipticals are closer to exponential (Binggeli,
Sandage \& Tarenghi 1984).  However, the evolution in profile shape could be a smooth
function of luminosity indicating similar formation and evolutionary scenarios for bright and dwarf
ellipticals (Jerjen \& Binggeli 1997).  And there is no reason to assume that dwarf
ellipticals are not `normal', in the sense of having peculiar structure.  They are
uniform as a class and we simply use the designation of `normal' to describe those
historically well studied, higher luminosity ellipticals.  Likewise, we maintain the
designation of dE to describe the dwarf ellipticals rather that the more common dSph
classification that extends to luminosities well below this study (Kormendy \etal
2009).

Investigating the connection between dwarf and normal ellipticals has been a
challenge due to the fact that normal ellipticals less luminous than $M_B = -18$ are
rare, but dE types (with luminosities down to $-$12) are numerous in nearby clusters
such as Virgo and Fornax.  This has led to a bifurcation in the samples with bright
ellipticals observed in the local Universe (distances out to 100 Mpc) and a large
sample of dwarf ellipticals from nearby clusters.  If there is a connection, a smooth
transition from bright to low luminosity ellipticals, then the nomenclature of
'dwarf' is simply an artificial slice by luminosity.  However, if there is a distinct
break in the structure of normals to dwarf ellipticals, then this may signal a
separate formation process or different evolutionary histories.

Importantly, as pointed out by Graham (2013), the non-linear behavior of correlations
between structural parameters and scaling laws extracted from fitting functions,
combined with a gap in elliptical samples with respect to luminosity (bright versus
dwarf) can result in an apparent dichotomy in the elliptical sequence.  While dwarf
ellipticals display a wide range in spatial and kinematic properties (Conselice \etal
2001) plus star population characteristics (Poggianti \etal 2001), a continuum with
respect to structure would establish a single scenario for most early-type galaxies
independent of their mass.  Thus, it is critical to fill the gap in luminosity and
extend the normal elliptical morphology sample to fainter luminosities with objects
that have well defined surface brightness profiles for a direct comparison of
structure.  That is the goal of this study.

\section{Sample \& Data Reduction}

This study is an outgrow from a study of bright ellipticals tracing colors from the near-UV
to the near-IR (Schombert 2016).  That sample was selected from the RSA and UGC
catalogs to find undisturbed ellipticals by morphology and isolation from bright
stars and other nearby galaxies.  For that sample, only 5\% of the galaxies were
fainter than $M_B = -19$.  To extend the elliptical sequence, we have selected 60
more ellipticals from the recent early-type catalog of Dabringhausen \& Fellhauer (2016)
specifically for low absolute magnitude.  Again, pure elliptical morphology and
isolation were the primary criteria, and the target had to be in the SDSS DR13 image
library.

In addition to faint ellipticals (classed as E in Dabringhausen \& Fellhauer), 52
dwarf ellipticals were also selected from the dwarf elliptical sample of Lisker,
Grebel \& Binggeli (2008) for study (again, isolation was the primary criteria).  Of
this 52, 49 are classed dE(N), eight are classed dE(nN) and five as dE(bc) based on
the Lisker scheme.  All of these galaxies are in the Virgo cluster with excellent
photometry from SDSS.  The Virgo sample was combined with a sample of group dE from
Dabringhausen \& Fellhauer for a total dwarf sample of 62 galaxies.  In addition, we
have supplemented the faint elliptical sample with 6 ellipticals from the ACSVCS
Virgo sample (Chen \etal 2010) plus 11 dwarf ellipticals from the Chen \etal study.
The combined sample (bright, faint and dwarf) contains 374 ellipticals with
photometry from SDSS $ugri$ images and the full data (luminosities, colors and
structural parameters) will be released in Schombert (2017).

\begin{figure}[!ht]
\centering
\includegraphics[scale=0.48,angle=0]{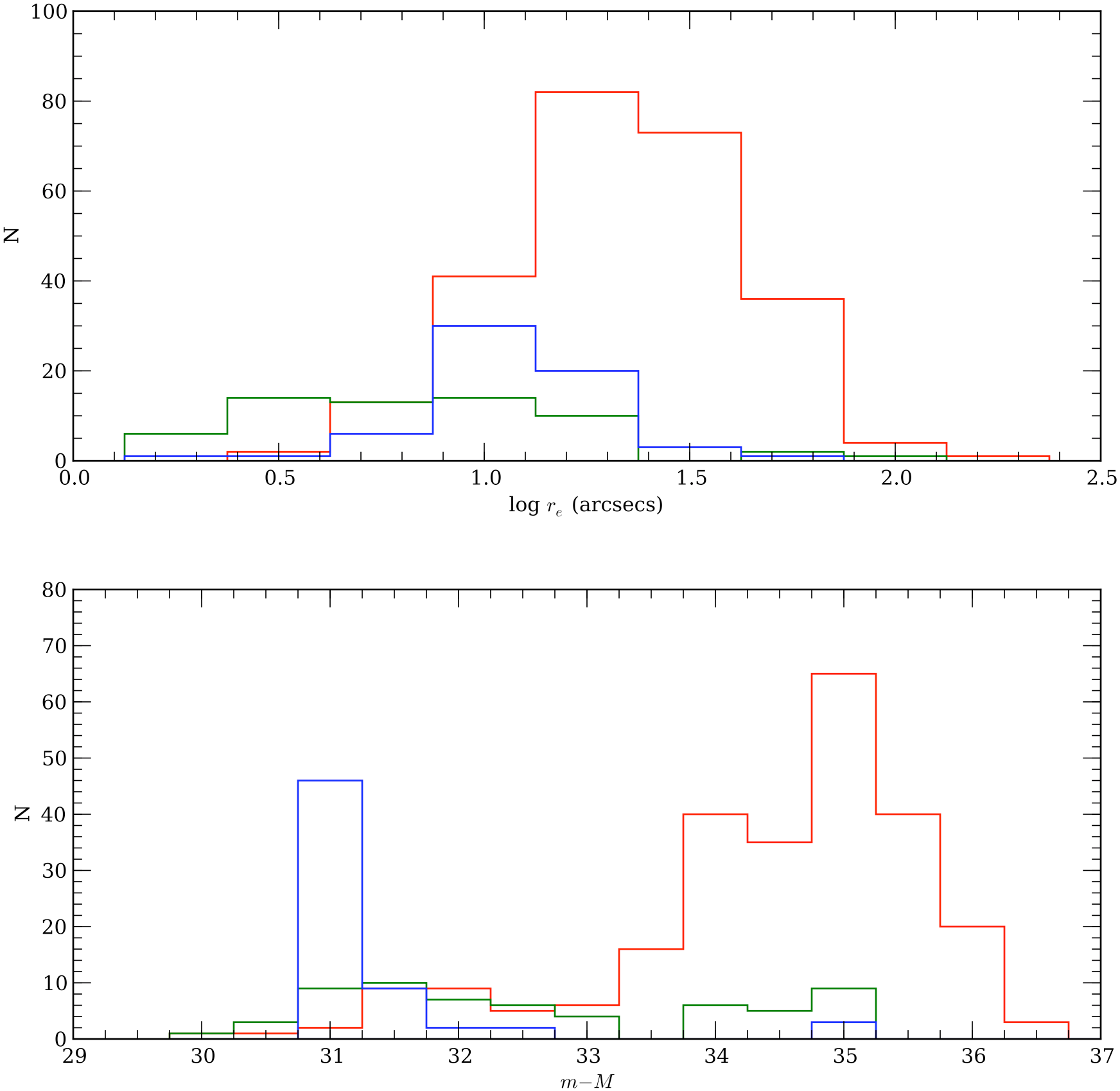}
\caption{\small Histograms of distance modulus and effective radius in arcsec units.
The red line is the bright ellipticals $M_g < -20$), green line is faint ellipticals
(M$_g > -20$) and the blue line is the dwarf elliptical sample.  The criteria for
isolation avoids many of the nearby rich clusters (e.g., Virgo and Fornax) for the
bright elliptical sample with a mean $m-M$ of 34.5.  The dE sample is concentrated in
the Virgo cluster at $m-M = 31.1$ and the faint ellipticals are distributed at
various distances in between.  The more distant bright elliptical sample means that
the scalelength structural parameters (such as effective radius, $r_e$) are similar
in arcsecs to the dE sample.  All the structural parameters are measured to be well
outside the radius where PSF effects dominate and at mean surface brightnesses well
above the noise limits of the sky brightness (see Schobmert 2017 for a more detailed
discussion of the data sample).
}
\label{dm_hist}
\end{figure}

For comparison, we have also extracted the structural parameters for 210 dE's from
Gavazzi \etal (2005), a deep Virgo study with the 2.4m Isaac Newton Telescope.  We
have 42 ellipticals in common with the Chen \etal sample and 27 with Gavazzi \etal,
reduced for their surface brightness profiles independently.  All the luminosities
and structure parameters were within 5\% of the photometric and fitting accuracies.
For galaxies in common, we have used our own photometry as we did not have access to
the raw surface photometry for the above published studies, and the remaining members
of their samples were too faint for SDSS analysis.

Data reduction of the flattened, calibrated images from the SDSS archive was performed
with the galaxy photometry package ARCHANGEL (Schombert 2011).  These routines, most
written in Python, have their origin back to disk galaxy photometry from the late
1980's and blend in with the GASP package from that era (Cawson 1987).  The package
has four core algorithms that 1) aggressively clean and mask images, 2) fit
elliptical isophotes to produce surface photometry, 3) repair masked regions then
perform elliptical aperture photometry and 4) determine aperture colors and
asymptotic magnitudes from curves of growth and determine accurate errors based on
image characteristics, such as the quality of the sky value.

The photometric analysis of ellipticals branches into four areas; 1) isophotal
analysis (the shape of the isophotes), 2) surface brightness determination and
fitting (2D images reduced to 1D luminosity profiles), 3) aperture luminosities
(typically using masked and repaired images and elliptical apertures) and 4)
asymptotic or total magnitudes (using curves of growth guided by surface brightness
data for the halos, see Schombert 2011).  Ellipticals are the simplest galaxies to
reduce from 2D images to 1D luminosity profiles since, to first order, they have
uniformly elliptical shaped isophotes (Jedrzejewski 1987).  Where many ellipticals
display disky or boxy isophotal shapes (Kormendy \& Bender 1996), this deviation is
at the few percent level and has a negligible effect on the surface brightness
profiles, aperture luminosities or colors values.  All surface brightness values are
determined using the generalized radius ($\sqrt{ab}$) rather than the major axis.
Using the major axis is only warranted if there is some confidence that the object is
oblate.  Dwarf ellipticals follow the same axial ratio distributions as bright
ellipticals (Lisker \etal 2007) so an oblate shape is not indicated for that sample
as well.

Fits to the surface brightness profiles followed the prescription of Schombert (2013)
for both the $r^{1/4}$ and S\'{e}rsic $r^{1/n}$ functions.  As noted in that study,
the value for the S\'{e}rsic $n$ index can vary depending on the region of the
profile with greater weight (typically the inner with its higher photometric
accuracy).  As the structural differences between normal and dwarf ellipticals
focuses on the behavior of the inner profile shape, we elect to use the inner profile
fitting procedure outlined in Schombert (2013).  This gives greater weight to the higher
surface brightness inner data points and uses photometric accuracy (RMS around each
elliptical isophote) to determine the weighting in the outer data points.

Fits were made to extinction corrected profiles.  While PSF corrections were applied,
all the galaxies were fit outside the 1.5 arcsec radius as an additional constraint
against PSF errors.  Surface brightness errors were determined from RMS errors on
each ellipse combined with error due to sky uncertainty.  The former dominates the
inner isophotes, the later dominates the outer data points.  Total magnitudes were
determined by two techniques, 1) fits to the curve of growth as an extrapolation to a
total magnitude and 2) a 20\% correction to the Kron magnitude determined from the
surface brightness profiles (see Schombert 2016).  Both techniques yielded the same
luminosity to within 2\% for 95\% of the sample.  Galaxies with unusual surface
brightness profiles, or curves of growth that did not converge, were eliminated from the
sample.

All distance related parameters used the CMB distance from NED (for the bright
ellipticals) or the distances found in the Dabringhausen \& Fellhauer catalog.  For
faint, nearby ellipticals, distances were also collected from redshift independent
distances found in NED (particularly important for the faintest of the normal
ellipticals, see \S3).  A majority of the dwarf ellipticals are in the Virgo cluster
and a distance modulus of 31.09 was assumed for all of them.

The distribution of distance moduli is shown in Figure \ref{dm_hist}.  The isolation
criteria produces a sample of bright ellipticals that avoids nearby rich clusters.  The
dwarf ellipticals are concentrated in the Virgo cluster.  In addition, the
distribution of effective radius (from S\'{e}rsic $r^{1/n}$ fits) is shown in units
of arcsecs to demonstrate that all the structural parameters are determined from
regions of the three samples that are similar in resolution element size and well
outside the PSF limited requiem.  Effective surface brightness for all three samples
were also well above the sky noise limits (see Schombert 2017 for a larger discussion
of the samples and analysis).

\section{Structural Relationships}

\begin{figure}[!ht]
\centering
\includegraphics[scale=0.48,angle=0]{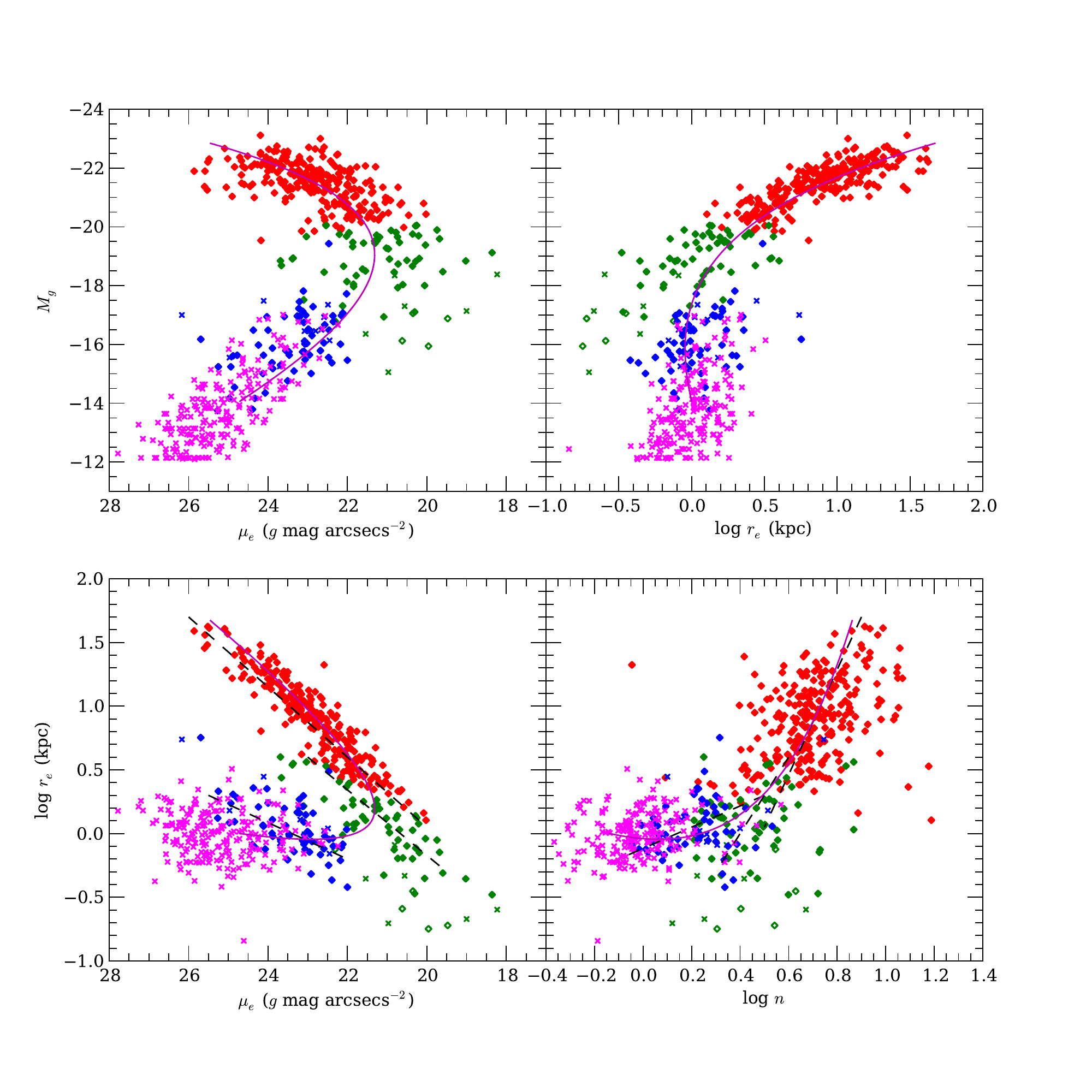}
\caption{\small Structural parameter space for SDSS $g$ magnitudes and surface
brightness fits.  The top two panels display total luminosity ($M_g$) versus
S\'{e}rsic effective radius ($r_e$) and surface brightness ($\mu_e$).  The bottom two
panels display the relationships between S\'{e}rsic parameters $r_e$, $\mu_e$ and the
concentration index $n$.  The normal elliptical sample (morphologically classed as
'E' and having power-law shaped profiles) are divided into bright (red) and faint
(green, the Chen \etal data is shown as crosses).  The dwarf elliptical sample
(morphologically classed as 'dE') are shown as blue symbols, the Gavazzi \etal dE
sample are shown as magenta symbols.  The curved relationships from Graham \&
Guzm\'{a}n (2003) are shown as magenta lines in each panel.  The open green symbols
are six faint ellipticals discussed in the text.  The black dashed lines are the
trends determined from template analysis in \S4 for each of the three subsets.
}
\label{str}
\end{figure}

Fits with the S\'{e}rsic function outputs three parameters, the effective radius
($r_e$, in kpc), the effective surface brightness ($\mu_e$, in $g$ mags
arcsec$^{-2}$) and the concentration or shape index $n$.  About 80\% of the sample
had $n$ values less than 6, which means that $r_e$ is approximately the half-light
radius ($r_h$) and $\mu_e$ is the surface brightness at the half-light point (Graham
\& Driver 2005).  All these structural parameters are correlated with stellar
luminosities (a proxy for total baryonic mass) as has been shown by Schombert (2013).
Two of these correlations are shown in the top panels of Figure \ref{str}.

Graham (2013) divides structural parameter behavior into three types 1) linear
relations, 2) non-linear or curved relations and 3) broken relations.  Foremost of
the linear relations is the luminosity-concentration correlation which is at the
heart of the photometric plane (Graham 2002) for the $n$ index can substitute for the
velocity dispersion in the usual Fundamental Plane correlations.  Whereas Schombert
(2013) found the $M_t$ versus log $n$ relation to be less well defined for bright
ellipticals, this is due to the degeneracy in the $n$ index for shallow profile
slopes typical to ellipticals brighter than $-$21.  In fact, as can be seen in Figure
\ref{str} by comparing the left panels, the $n$ index varies uniformly with luminosity
and the new faint elliptical sample connects the dwarf sequence to the bright
ellipticals in a smooth, even fashion.  The slope for that relationship is consistent
with the slope found by Graham \& Guzm\'{a}n (2003) for the full sample from dwarf to
bright ellipticals.

The relationships comparing effective radius and surface brightness to total
luminosity are displayed in the top panels of Figure \ref{str}.  The total magnitude
for the sample is shown on the y-axis, $M_g$, plotted against effective radius
($r_e$) and surface brightness ($\mu_e$).  The sample is divided into three subsets,
bright ellipticals (with $M_g < -20$, red), the faint ellipticals (green) and the
morphologically classified dE's (blue).  The faint ellipticals all have "E"
classifications, confirmed by visual inspection of their images and comparison of
their surface brightness profiles to brighter ellipticals.  None have the diffuse
appearance that distinguishes the dE class.

Several well-known historical trends are displayed.  The trend of decreasing
effective surface luminosity with increasing total luminosity for normal ellipticals
is evident.  As outlined in Schombert (2013), this reflects the increasing
shallowness of the typical elliptical profile slope as the galaxy grows larger and
brighter.   There is also the obvious size-luminosity relation (Kormendy 2009) for
bright ellipticals in the top right panel of increasing effective radius ($r_e$) with
elliptical total luminosity.

While both luminosity relations for bright ellipticals (red) are roughly linear,
there is significant curvature as anticipated by analysis of Graham \& Guzm\'{a}n (2003).
As pointed out by Graham (2013), the lack of strict homology for ellipticals leads to
non-linear relationships between the various structure parameters extracted from
S\'{e}rsic fits.  The curved relationships, as outlined in Graham (2013), are defined
from the fact that the central surface brightness, $\mu_e$, and the concentration
index, $n$, are linearly correlated with total luminosity.  One can then derive the
expected correlations between effective radius ($r_e$) and surface brightness
($\mu_e$) versus total luminosity based on the their coupling within the S\'{e}rsic
function.  This will only apply if all ellipticals are well fit by the S\'{e}rsic
function, but this is certainly the case if one ignores the complications introduced
by core processes which are not relevant to this discussion.

The resulting curved relations from Graham (2013) are shown in Figure \ref{str} as
the solid lines.  While the correlation between $\mu_o$ and $n$ breaks down for the
very brightest ellipticals (there is a degeneracy in the fitting process, see
Schombert 2013), the curved relations are an excellent description of the behavior of
the fitting parameters are a function of total luminosity (top panels).  The same
curved relations are plotted in the fitting parameter space (bottom panels) which are
also in excellent agreement, connecting the bright and dwarf ellipticals.  These
revised structure correlations are the main argument against a dichotomy between
normal and dwarf ellipticals as the curved relation connect the two branches into one
branch, a continuum of ellipticals by luminosity that follow the same fitting
function (a S\'{e}rsic r$^{1/n}$ function).

Also, historically, the claim for a dichotomy focused on fitting linear relations to
the brightest ellipticals which clearly placed the dwarf ellipticals in a separate
part of structural parameter space (Kormendy 1977).  For example, the $r_e$ versus
total luminosity diagram (top right panel in Figure \ref{str}, often called the
size-luminosity diagram) displays a nearly linear behavior for ellipticals brighter
than $-$19 (the typical luminosity cutoff for a dwarf elliptical).  The
morphologically classed dwarf ellipticals display larger $r_e$ for their luminosity
(in agreement with their diffuse appearance) with little indication of a strict
correlation with luminosity itself.  Other studies have attempted to fit two separate
linear relations to the normal and dwarf ellipticals (see Dabringhausen \etal 2008,
Lisker 2009), but, in fact, the dE's effective sizes seems uncorrelated
(although always between 0.5 and 2 kpcs) with respect to their total stellar mass,
rather than a proper linear relationship.  Although it is true that all dE's with
$r_e$ larger than 1 kpc have a mean $M_g$ of $-$16.5 and the smaller subset has a
mean $M_g$ of $-$15.8, barely significant.

The relationship between $r_e$ and $\mu_e$ (the Kormendy relation, bottom left panel
in Figure \ref{str}) displays the most salient characteristic differences between
normal and dwarf ellipticals.  The trend for fainter effective surface brightness
($\mu_e$) with increasing size ($r_e$) is evident.  The degree of linearity is
questionable but, to first order, the extension of the bright elliptical sequence to
fainter luminosities is roughly linear (a statement of the accuracy of the S\'{e}rsic
fitting function over a large range in luminosities).  The curved relation from
Graham (2013) is shown and connects the normal to the dwarf sequence, but at the
expense of ignoring the fainter ellipticals.  The size-surface brightness relation
also predicts nearly constant effective radius ($r_e$) for the dwarf ellipticals
(thus, decreasing surface brightness results in a more diffuse appearance with
decreasing luminosity).  The trend for this sample is lowest $\mu_e$ dwarf
ellipticals are slightly larger in $r_e$ compared to the brighter dE's.  Most
importantly, the new sample of fainter normal ellipticals do not follow the curved
relationship, although they are well fit by the S\'{e}rsic function (see below).

Key to the claim of a dichotomy between normal and dwarf ellipticals are the very
faintest normal ellipticals in Figure \ref{str}, those with small $r_e$ but with high
surface brightnesses ($\mu_e < 21\, g$ mag arcsecs$^{-2}$).  They appear to extend the
bright elliptical sequence to smaller $r_e$ and brighter $\mu_e$ in a roughly linear
fashion from the bright ellipticals.  In particular, they occupy portions of
structural space that are outside the predications from the Graham curved relations
that are intended to connect the normal and dwarf sequences.  However, Chen \etal
(2010) argues that using inaccurate morphological information artificially forces a
dichotomy by dividing the elliptical sample exactly where the curved relations
connect the bright and faint elliptical sequence.  It is certainly true that the
number of ellipticals between the top of the dE sequence (at $M_g = -18$) and the
bottom of the bright elliptical (at $M_g = -19$) is quite sparse even when combining
Chen \etal and our sample (only 34 galaxies not classified as dE are less then $-$19
in luminosity).  Chen \etal find only four galaxies in higher surface brightness
region (marked as green crosses in Figure \ref{str}) and conclude that secular
processes (such as tidal stripping, Bekki \etal 2001) could have produced their unusual
structure in the very rich Virgo environment.

\begin{figure}[!ht]
\centering
\includegraphics[scale=0.48,angle=0]{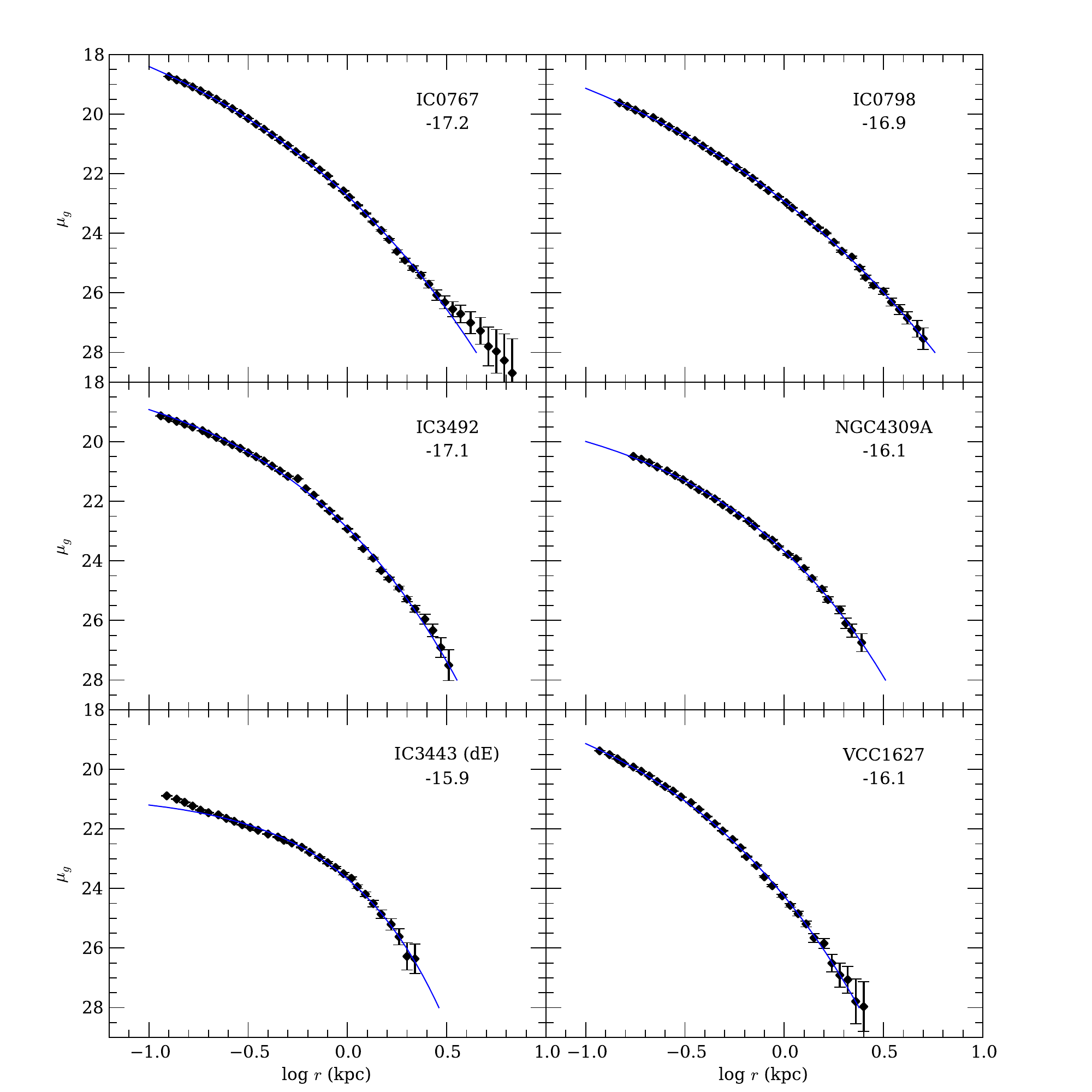}
\caption{\small Surface brightness profiles for five normal ellipticals with
luminosities less than $-$17.5 (shown as open symbols in Figure \ref{str}).  Blue
lines display their best $\chi^2$ S\'{e}rsic fits.  All six are well fit by the
S\'{e}rsic function.  IC3443 (bottom left panel) is a classic dE of similar
luminosity shown for comparison.  Both IC3443 and VCC1627's greyscale images is shown in Figure \ref{compare}.
}
\label{special}
\end{figure}

From the
comparison with the Chen \etal study, it is clear that any claim of dichotomy is based
whether the ellipticals, classified as E, between $-$19 and $-$16 in Figure \ref{str}
are actually distinct from the dE galaxies of similar luminosity.  Chen \etal
concludes that this inflection point is due to structural non-homology (presumingly
from the homologous dE's to the weak homology found for bright ellipticals).  And a
lack of dichotomy is supported by the smooth transition in other characteristics from
dE to E (e.g., color and metallicity).  

To examine this interpretation more closely, five of these high surface brightness,
faint ellipticals are marked by open symbols in Figure \ref{str} and their individual
profiles are shown in Figure \ref{special} (along with a dE of similar luminosity,
IC3443).  All five are well fit by the S\'{e}rsic function with low effective radii
($r_e$) and bright effective surface brightnesses ($\mu_e$).  All five profiles
follow the trend in profile shape displayed by the bright end of the normal
elliptical template profiles, meaning they are more power-law shaped than exponential
(i.e., $n > 1$, see \S4).

\begin{figure*}[!ht]
\centering
\includegraphics[scale=1.2,angle=0]{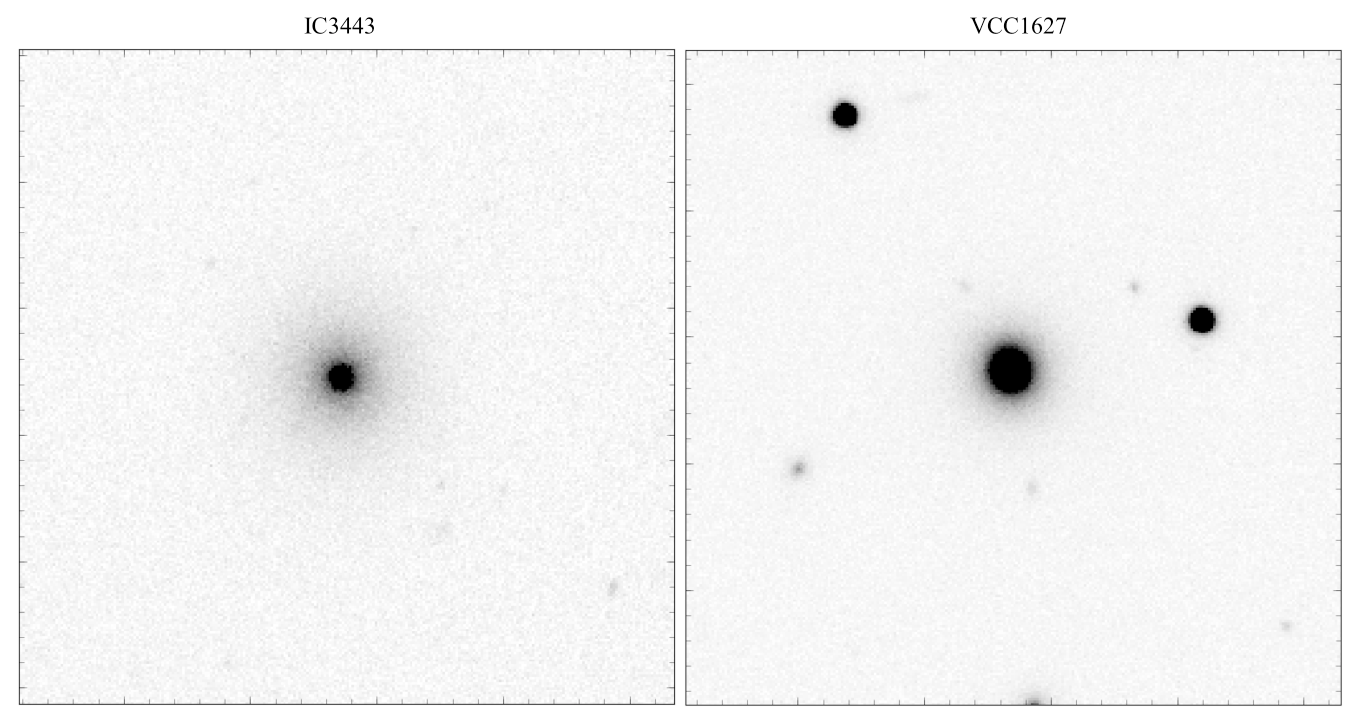}
\caption{\small SDSS $g$ images for the dE IC3443 ($M_g = -15.8$) and E class VCC1627
($M_g = -16.1$).  Each frame is 100 arcsecs on a side (approximately 8 kpc for each
galaxy) where the greyscale is set at 23 $g$ mag arcsecs$^{-2}$ for the blackest
level and 26 $g$ mag arcsecs$^{-2}$ for the sky level in both frames.  The diffuse
appearance of IC3443 is the visual signature of the dE class compare to normal
ellipticals.
}
\label{compare}
\end{figure*}

In addition, their visual morphology is clearly distinct from dwarf ellipticals.
Figure \ref{compare} displays a low contrast greyscale of a visually classified dwarf
elliptical (IC3443, type dE(N) from Lisker \etal 2008) side-by-side with a faint
normal elliptical (VCC1627, whose profile is in Figure \ref{special}).  Each galaxy
was selected to be similar in ellipticity and luminosity.  The diffuse appearance to
IC3443 is the quality that morphologists use to definite the dE class and is obvious
in the greyscale image (see also, Figure 3 from Wirth \& Gallagher 1984).  This
distinction is important, as both galaxies have similar stellar masses and the
difference relates to individual galaxy's profile slope, where the more shallow
profile produces a more diffuse appearance.

Distance errors might account for some of the faint ellipticals deviating from the
dwarf elliptical sequence.  For example, under estimating their distance would produce
larger effective radii ($r_e$) and move them onto the dE sequence in the top right
panel of Figure \ref{str}.  However, surface brightness is distance independent
(aside from redshift corrections) and no distance errors can move the faintest
ellipticals to fainter effective surface brightnesses ($\mu_e$) in the top left panel of
Figure \ref{str}.

The faint elliptical region in structure diagrams are known to be populated by the
class of elliptials called compact ellipticals (cE).  The prototype object being M32
with the largest sample of cE type ellipticals from the AIMSS project (Norris \etal 2014).
Based on their Figures 11 and 14, the AIMSS cE's would overlap the smallest ($r_e$
between 0.1 and 1 kpc) and highest surface brightness ($\mu_e$ greater than 21) faint
ellipticals from our sample.  The AIMSS survey extends the elliptical sequence beyond
the cE and faint ellipticals to luminosities given by ultra-compact dwarfs (UCD, $M_g
> -12$).  Notability, the sequence of cE and UCD's connects linearly with the
brighter ellipticals in terms of scalelength.  The relationship between surface
brightness of cE's and normal ellipticals increases with decreasing luminosity, but
the UCD's decrease in effective surface brightness with decreasing luminosity in the
same fashion as the dwarf ellipticals.

Comparison with the cE sample is problematic as a majority of the AIMSS cE's are
embedded in the envelopes of other bright galaxies or in the center of high density
clusters making extraction of their surface brightness profiles nearly impossible.
For a handful of AIMSS cE's with large enough distances from other galaxies for
adequate surface brightness analysis, their luminosity and structure characteristics
were identical to the faint ellipticals of our sample with similar luminosities.
In particular, their effective surface brightnesses were high and their
characteristic scalelengths were low in agreement with their low luminosities.

It has been proposed by many studies (Bekki \etal 2001, Chilingarian \etal 2009,
Pfeffer \& Baumgardt 2013) that compact ellipticals are formed by tidal stripping of
either brighter ellipticals or disk galaxies (removing the disk and leaving a compact
bulge).  In addition, tidally induced star formation can increase the stellar
densities in the core regions, increasing the inner surface brightness of these objects.
Just tidally stripping by itself will not change the shape of the inner surface
brightness profile of a bulge or elliptical.  However, the effective surface
brightness is approximately the luminosity of a galaxy inside its effective radius,
divided by the area given by the effective radius.  Thus, reducing the effective radius
by tidally stripping will increase the effective surface brightness, even if not
directly changing the stellar densities in the core regions of a galaxy.  Thus, the
argument is made that ellipticals occupying the region given by a linear
extrapolation of the bright elliptical sequence (high $\mu_e$, low $r_e$ and
intermediate $n$ values) are, in fact, tidally stripped bright ellipticals and not
part of the structural sequence from E to dE imposed by formation processes.

Of the 62 faint ellipticals in our sample, slightly less than half of them are
members of groups or have close companions.  Those which have companions, they are
all the lesser member of the pair (statistically, this is expected).  Tidal effects
could strip the outer envelopes of many of these faint ellipticals, reducing $r_e$.
The key issue is if the profile shape of faintest ellipticals have
more in common with the bright elliptical sequence or the dwarf sequence.  

We conclude the analysis from structure relations by noted that Graham \& Guzm\'{a}n
(2003) argue that the curved relations demonstrate that the apparent dichotomy is, in
fact, due solely to a smooth and steady change in profile shape from bright
ellipticals to faint dwarfs.  The linear portions of the structure diagrams occur
where the concentration index, $n$, is large and effective radius, $r_e$, is more
strongly correlated with luminosity.  The underlying linear correlations are between
luminosity and $n$ or $\mu_e$, which are more critical in defining a galaxy's profile
shape.  Thus, there is no dichotomy as the various structure diagrams (with seemingly
separate regions of parameter space) are simply reflecting a smooth transition from
bright to dwarf ellipticals in profile shape.  

However, as noted by Janz \& Lisker (2009), the first order behavior of the various
structure diagrams is explained by varying profile shapes, but there is a significant
number of ellipticals brighter than dE's with smaller effective radius ($r_e$).
As noted above, these objects could be the result of tidal stripping, as proposed for M32 (Bekki
\etal 2001), but they conclude that distribution of structural values is much
larger than that expected by random scatter and that, in particular, the
size-luminosity relation can not be fully explained by a truncated profile shape.  A
averaged comparison of profile shapes requires the use of template analysis pioneered
by Schombert (1987), and will be explored in the next section.

\section{Template Analysis}

A clearer view of the elliptical dichotomy can be deduced through the use of template
analysis.  This technique, first used for brightest cluster ellipticals (Schombert
1987), and expanded for 2MASS ellipticals in Schombert (2013), uses the total surface
brightness profile of a galaxy to construct average templates, rather than forcing a
fitting function onto a profile shape and then extracting structure parameters from
the function best fits.  Template comparison is non-parametric and has the advantage
of being free of coupling effects between fitting function parameters plus presents a
more accurate measure of whether isophotal structure is a smooth function of total
galaxy mass (a key test of homology from various galaxy formation scenarios).  It has
the disadvantage of not reducing a 2D profile into a few 1D parameters which makes
comparison of large samples difficult.

To determine if there is a dichotomy between dwarf and normal ellipticals in terms of
structure, we are basically asking if dwarf and normal elliptical surface brightness
profiles are self-similar.  In other words, are dwarf ellipticals simply scaled down
versions of their brighter (i.e. more massive) cousins, as the curved relations from
Graham (2013) suggest.  The problem with this determination is that normal
ellipticals themselves do not display absolute structural homology (Schombert 2015).
Their profile shapes do change smoothly with luminosity (or scalelength, so-called
weak homology), but they are not self-similar as can be seen in Figure 3 of Schombert
(2015).  The brightest ellipticals are mostly r$^{1/4}$ in shape, but gradually
develop more curvature (a lower S\'{e}rsic $n$ index) with decreasing luminosity.
Thus, while normal ellipticals are homologous in profile shape within limited luminosity
bins, they do not strictly display complete homology.

\begin{figure}[!ht]
\centering
\includegraphics[scale=0.48,angle=0]{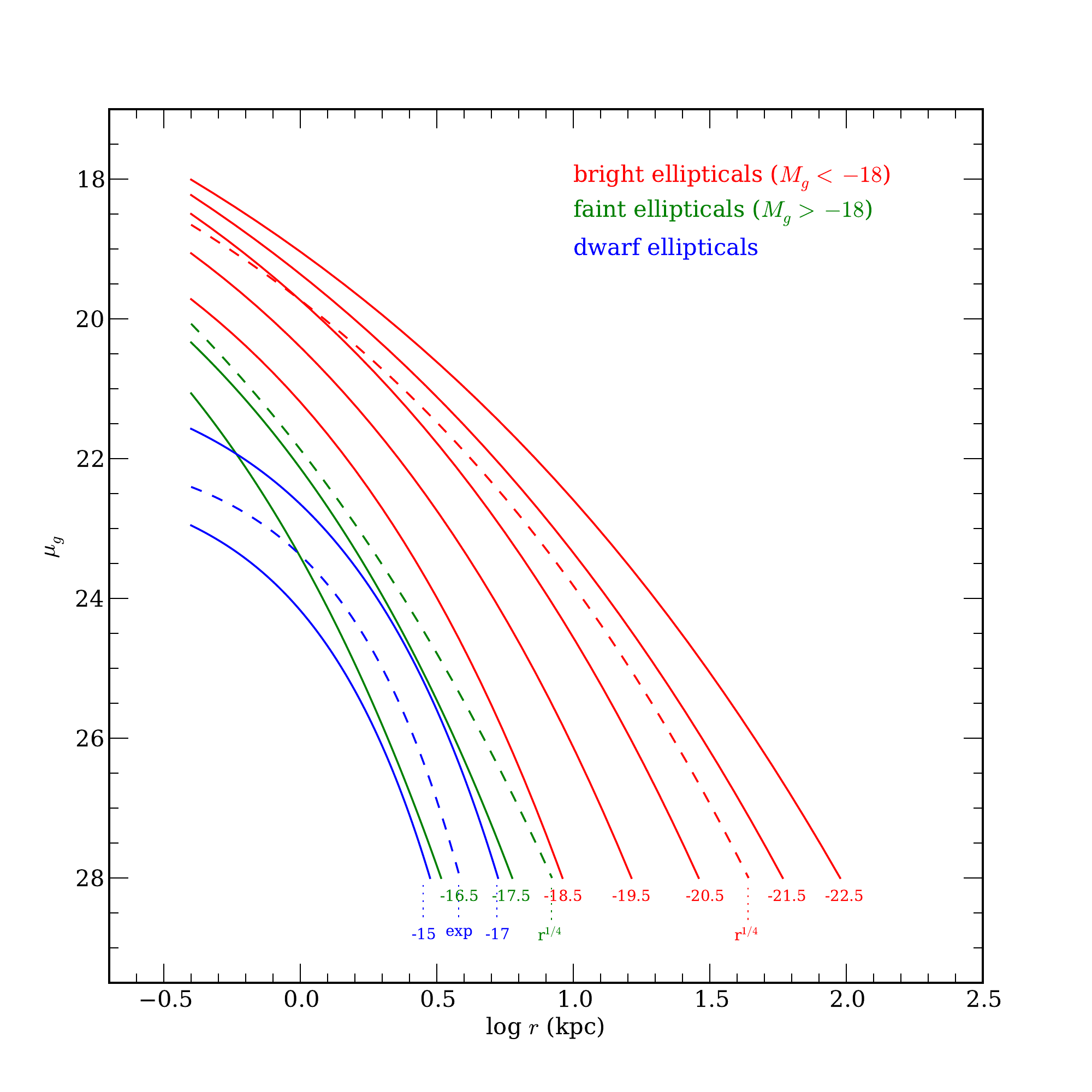}
\caption{\small Template profiles constructed by the methods outlines in Schombert
(2013).  The red profiles (parameterized by total magnitude) are for bright ellipticals,
and agree with the $V$ templates from Schombert (1984) and the 2MASS $J$ profiles.
While the profiles as a function of luminosity are not self-similar (homology), the
change with luminosity is smooth and quantifiable.  The green templates are
constructed from faint normal ellipticals and follow the same profile trend as the
bright ellipticals.  The templates for dwarf ellipticals are shown in blue and are
clearly distinct from the normal elliptical templates with lower central surface
brightness and more extended isophotal radii.  Reference profiles displaying the
r$^{1/4}$ and exponential shape are shown as dashed lines.
}
\label{sfb}
\end{figure}

On the other hand, this uniform change in structure with luminosity can be used for
comparison to dwarf ellipticals.  For the new sample of faint ellipticals can be
compared both to the morphologically classified dwarf ellipticals and the brighter
ellipticals with the same template analysis technique.  The results from this
analysis are shown in Figure \ref{sfb}.  As defined in \S2, the normal elliptical
sample is divided into two samples, bright and faint with a luminosity cut-off at
$M_g = -20$).  The bright sample contains 252 galaxies, the faint sample contains 60
galaxies (green symbols in Figure \ref{str}).  Template averaging used 1/2 mag bins
from $-$23 to $-$16.  For comparison, the lowest luminosity template for the 2MASS
$J$ templates were $-$21.5 $J$ which corresponds to roughly $-$20 $g$.

The resulting normal elliptical templates agree well with the templates defined in
Schombert (2013) using 2MASS $J$ profiles.  A similar shift from r$^{1/4}$-like to
significantly more curvature with decreasing luminosity is evident (for reference,
two pure r$^{1/4}$ profiles are indicated in Figure \ref{sfb}).  The faintest
luminosity bins (below $-$18) are shown in green as these luminosities are equivalent
to the luminosities of the dwarf elliptical sample.  A luminosity cutoff at $-$18
seems arbitrary but, in fact, there are very few dwarf ellipticals (by morphology)
with higher luminosities.  We note that the faintest normal elliptical profiles
follow the exact same trend as outlined by the brighter profiles (i.e., they continue
the weak homology trends).  This would indicate that the continuation of the linear
trends in the structure diagrams of Figure \ref{str} are accurate and the normal
ellipticals follow a different structural sequence compared to dwarf ellipticals.
The S\'{e}rsic fits to the templates are shown in Figure \ref{str} as dashed lines.

We perform the same analysis on the 62 dwarf ellipticals in the sample binned into
four luminosity bins starting at $-$14.  Two of the templates ($-$15 and $-$17) are
shown for reference in Figure \ref{sfb} (blue curves).  It is immediately obvious
that the dE templates have a very different shape compared to the normal ellipticals
and are {\it not} a continuation of the normal elliptical sequence.  For a set value
of galaxy luminosity (i.e., stellar mass), the dE templates are lower in central
surface brightness, more extended in isophotal radius and display greater curvature
(i.e., lower $n$ values) than the normal ellipticals.  Drawn for reference in Figure
\ref{sfb} is an exponential ($n=1$) profile (dashed blue line).  While the normal
ellipticals are nearly r$^{1/4}$ (although deviating systematically at lower
luminosities), the dwarf ellipticals are closer to an exponential profile than
$r^{1/4}$ (something known for many decades, see Caldwell 1983; Binggeli \& Cameron
1991).

Note the dE templates have greater isophotal radii than normal ellipticals at any
luminosity, and are offset by lower central surface brightness.  This reflects the
qualitative diffuse appearance determined by visual morphology and seen in Figure
\ref{compare}.  Interestingly, the dwarf ellipticals display greater homology than
the normal ellipticals.  On average, dwarf ellipticals are more self-similar than
normal ellipticals and can be scaled up and down in effective radius to reproduce the
full range in luminosity displayed by the sample, although that range in total
luminosity is much smaller than the normal elliptical sample, detailed inclusion of
fainter dE's would quantify this statement.

The class of compact ellipticals (cE) also follow the normal elliptical sequence.  Although it
does not seem to have been noticed by previous studies, all the compact ellipticals have
luminosities less than $-$20 $M_g$.  Presumingly a tidal origin to this class of
ellipticals would naturally produce a smaller, fainter object.  Their close
association with bright companions supports this hypothesis, although, statistically,
faint ellipticals would be the lesser companion to a pair or small group.  Thus,
their smaller effective radii ($r_e$) are expected for the smaller profiles per given
luminosity.  There is nothing peculiar about their profile shapes, although many have
poorer accuracy at low surface brightness levels (due to isophotal confusion with their
brighter companions) and may display undetected tidal stripping signatures in their outer
isophotes.  However, to within a few $r_e$, their profile shapes and characteristic
surface brightnesses are well matched to the templates and by an extrapolation from
brighter, isolated ellipticals.

\section{Conclusions}

Given the wide dispersion in kinematics for bright ellipticals, presumingly from a
history of mergers (Bender \etal 1992), it is surprising that their structure is as
uniform as given by Figure \ref{str} or as indicated by template construction.  Thus,
a structural dichotomy between normal and dwarf ellipticals also seems odd in
comparison to many numerous astronomical relationships that smoothly trace the
ellipticals sequence over 12 magnitudes in luminosity.  For example, color and
metallicity progresses smoothly from dwarf to normal ellipticals (driven by stellar
mass, as it relates to the termination of galactic winds, Poggianti \etal 2001).
Even some structural parameters display no break with luminosity, such as the
concentration index $n$ and central surface brightness ($\mu_o$).  In fact, key to
the discussion of dichotomy is the combined behavior of the $n$ index with luminosity
and size (Graham \& Guzm\'{a}n 2003; Gavazzi \etal 2005).  For the $n$ index is a
powerful tool in parameterizing ellipticals and substitutes for velocity dispersion
to produce a 'Photometric Plane' for ellipticals comparable to the Fundamental Plane
(Graham 2002).  The $n$ index also links dwarf and normal ellipticals by varying
smoothly, and linearly, with luminosity.

Over time, two schools of thought have formed with respect to the E-dE dichotomy.
One school argues that dwarf and bright ellipticals represent one structural family
with a gradual increase in $n$ with luminosity (Jerjen \& Binggeli 1997; Graham \&
Guzm\'{a}n 2003) and, presumably, a common origin scenario.  The other school insists
that dwarfs and bright ellipticals are structurally distinct regardless of the
scaling relationships presented from S\'{e}rsic fits due to a separation seen in
numerous physical properties between the two types of ellipticals (Kormendy \etal
2009; Janz \& Lisker 2009) particularly their kinematic separation in the Fundamental
Plane (Bender \etal 1992).  Chen \etal (2010) argues that one should ignore
morphological classifications (as they are subjective) and direct our analysis to the
various fitting parameters.  Under this scheme, a vast majority of ellipticals follow
one sequence with a small minority of objects displaying deviant structure (such as
high central surface brightnesses) that could well be the result of environmental
factors.

Overall, there is no strict boundary by mass, size, density or kinematics to define a
dwarf elliptical from a normal elliptical (aside from an artificial luminosity
division deduced from morphology), but structure seems to be the singular feature.
Certainly structure, as it reflects into visual appearance, is the primary
consideration that morphologists use to divide ellipticals into dwarfs and normal or
compact classes.  This is also true for the late-type galaxies, such as dwarf
irregulars and disk galaxies, which separate by scalelength even though both types
are well fit by an exponential profile (Schombert 2006).  This reinforces a
connection between dI's and dE's (Grebel \etal 2003) as structural differences
usually signal varying formation scenarios (Driver \etal 2011) or strong merger
histories to disrupt the original structural form.

In summary, the evidence presented in this study supports the dichotomy in structure
between normal and dwarf ellipticals despite the wide number of non-structural
correlations that argue for a continuum between elliptical types.  The evidence falls
into three parts.  First, the extension of the normal elliptical sequence to fainter
luminosity extends to the linear relations beyond the expectations from S\'{e}rsic
fits that has been interpreted as connection between normal and dwarf ellipticals.
Second, many of those faint ellipticals have structural properties (such as effective
surface brightness) that are well outside the range of dwarf ellipticals of the same
luminosity (i.e., stellar mass).  A sufficient number of these ellipticals are found
over a range of environments to dismiss the conjecture that this is a small subset
whose structure has been disturbed by environmental effects.  

Third, and the most salient point, the template analysis clearly demonstrates that
the average profiles of dwarf ellipticals are distinct from the normal elliptical
sequence of profile shape with luminosity.  This last result derives directly from
the conclusion that both dwarf and normal ellipticals display, at least, weak
homology.  Bright ellipticals are clearly not self-similar (Schombert 2015), but do
display an quantifiable change in profile shape with luminosity.  Compact or faint
ellipticals also follow the bright elliptical template sequence and are distinct from
the shape of dE's.  Dwarf ellipticals display stronger homology (you can roughly
scale any dE profile upward or downward in luminosity as they are, on average,
exponential in shape).  But, dwarf ellipticals profiles can not be scaled into normal
ellipticals without significant changes to their relationship between characteristic
surface brightness and radius.  The two types of ellipticals appear to follow
separate evolutionary histories with respect to structure.  As dE's are only found in
clusters, it is possible that some environmental process dominates their structural
formation and evolution.

A separate structural path for dE's would argue for the parallel galaxy sequence proposed
by Kormendy \& Bender (2012).  In this scheme, ellipticals, S0's and gas-rich disks
form parallel sequences defined by formation processes and environmental secular
evolution can reshape their general appearance.  Structurally, dE and SO's are more
closely related than normal ellipticals and dE's.  This would make the dE sequence the
low luminosity counterparts to the higher luminosity S0's, although their kinematics
differ.

\noindent Acknowledgements: 

The software and funding for this project was supported by NASA's Applied Information
Systems Research (AISR) and Astrophysics Data Analysis Program (ADAP) programs.  Data
used for this study was based on observations made with SDSS where funding has been
provided by the Alfred P.  Sloan Foundation, the Participating Institutions, the
National Science Foundation, the U.S. Department of Energy, the National Aeronautics
and Space Administration, the Japanese Monbukagakusho, the Max Planck Society, and
the Higher Education Funding Council for England, In addition, this research has made
use of the NASA/IPAC Extragalactic Database (NED) which is operated by the Jet
Propulsion Laboratory, California Institute of Technology, under contract with the
National Aeronautics and Space Administration. 

\pagebreak

\end{document}